\documentclass[12pt]{article}
\newcommand{\ba}{\begin{array}}
\newcommand{\ea}{\end{array}}

\newcommand{\nn}{\nonumber\\}

\newcommand{\C}{{\bf C}}

\newcommand{\del}{\partial}

\newcommand{\rar}{\rightarrow}

\newcommand{\fr}{\frac}

\newcommand{\scr}{\scriptsize}

\newcommand{\at}{\tilde{a}}
\newcommand{\bt}{\tilde{b}}
\newcommand{\ct}{\tilde{c}}
\newcommand{\dt}{\tilde{d}}

\newcommand{\zt}{\tilde{z}}
\newcommand{\wt}{\tilde{w}}
\newcommand{\delt}{\tilde{\partial}}

\makeatletter
\@addtoreset{equation}{section}

\makeatother
\usepackage{pmat}
\usepackage{amsmath,amsfonts,amssymb}
\usepackage{graphicx}

\begin{document}

\begin{titlepage}

\null
\begin{flushright}
%arXiv:1101.0005\\
%August, 2013
\end{flushright}

\vskip 2cm
\begin{center}

  {\Large \bf Noncommutative Solitons and Quasideterminants}

\vskip 2cm
\normalsize

  {\large Masashi Hamanaka}\footnote{
E-mail: hamamnaka@math.nagoya-u.ac.jp}

\vskip 2cm

  {\it Nagoya University, Department of Mathematics,\\
              Chikusa-ku, Nagoya, 464-8602, JAPAN}

\vspace{10mm}

\end{center}

\vskip 10mm

%\vspace{10mm}

\begin{center}

\vspace{1mm}

{\bf{Abstract}}

\end{center}

We discuss an extension of soliton theory and integrable systems
to noncommutative spaces, focusing on integrable aspects of
noncommutative anti-self-dual Yang-Mills equations. We give 
a wide class of exact solutions by solving a Riemann-Hilbert problem
for the Atiyah-Ward ansatz and present B\"acklund transformations
for the $G=U(2)$ noncommutative anti-self-dual Yang-Mills equations. 
We find that one kind of noncommutative determinant, quasideterminants, 
play crucial roles in the construction of noncommutative solutions. 
We also discuss reduction of a noncommutative 
anti-self-dual Yang-Mills equation to noncommutative integrable
equations. This is partially based on a collaboration 
with C.~R.~Gilson and J.~J.~C.~Nimmo (Glasgow).

\end{titlepage}

\newpage

\section{Introduction}

The extension of integrable systems and soliton theories
to non-commutative (NC) space-times
\footnote{In the present paper, the word ``noncommutative'' always
refers to generalization to noncommutative spaces,
not to non-abelian and so on.}
has been studied by many authors over the past couple of years and 
various kinds of integrable-like properties
have been revealed. 
(For reviews, see \cite{DiMH_proc}%, Hamanaka_proc, Hamanaka_proc2, 
%Kupershmidt, Lechtenfeld, Mazzanti,
 - \cite{Tamassia}.)
This is partially motivated by the recent developments
of noncommutative gauge theories on D-branes. 
In the noncommutative gauge theories, the 
noncommutative extension corresponds to
the presence of background flux, and
in the effective theory of D-branes, noncommutative solitons 
can be identified with the lower-dimensional D-branes. 
(For reviews, see e.g. \cite{DoNe}%, Hamanaka_PhD, Harvey, 
 - \cite{Szabo}.)
This makes it possible to reveal some aspects
of D-brane dynamics, such as tachyon condensations, 
by constructing exact noncommutative solitons and studying their properties.

%Hence exact analysis of noncommutative solitons just leads to
%that of D-branes and various applications to D-brane dynamics
%have been done successfully. 
%In this sense, noncommutative solitons play important roles
%in noncommutative gauge theories.

Most of noncommutative integrable equations such as noncommutative 
Korteweg-de Vries (KdV) equations belong, 
apparently, not to gauge theories, but to scalar theories.
However, it has now been proved that they can be derived {}from 
noncommutative anti-self-dual (ASD) Yang-Mills (YM) equations
by reduction (e.g. \cite{Hamanaka06_NPB, Hamanaka05_PLB}),
which was first conjectured explicitly
by the author and K.~Toda \cite{HaTo}. (The original commutative one
was proposed by R.~Ward \cite{Ward}
and hence this conjecture is sometimes
called the {\it noncommutative Ward's conjecture}.)
As noncommutative anti-self-dual Yang-Mills equations
belong to gauge theories,
% and hence 
lower-dimensional many integrable equations 
%which apparently belong to scalar theories 
must have physical correspondence (in the background flux), and
therefore analysis of exact soliton solutions of noncommutative
integrable equations could be applied to
the corresponding physical situations
in the framework of $N=2$ string theories \cite{Hull} - \cite{OoVa}.
(In this context, the signature is not Euclidean $(+++\,+)$ 
but ultrahyperbolic $(++-\,-)$ and 
the $N=2$ string theory lives in this signature.)

Furthermore, integrable aspects of anti-self-dual Yang-Mills equation 
can be understood {}from the geometrical framework of
the {\it twistor theory}. Via the Ward's conjecture,
the twistor theory gives a new geometrical viewpoint
into the lower-dimensional integrable equations
and some classification can be carried out in such a way.
These results are summarized in the book of Mason
and Woodhouse elegantly \cite{MaWo}. (See also \cite{Dunajski, Dunajski2}.)

In this paper, we discuss integrable aspects of the
noncommutative anti-self-dual Yang-Mills equations 
{}from the viewpoint of the noncommutative twistor theory. 
We give a series of noncommutative Atiyah-Ward ansatz solutions 
by solving a noncommutative version of the Riemann-Hilbert problem. 
The solutions include not only noncommutative instantons (with finite action) 
but also noncommutative non-linear plane waves and so on (with infinite action). 
We also find that noncommutative determinants of
a particular kind, the quasideterminants, % (for a good survey, see \cite{GGRW}), 
play crucial roles in construction of exact solutions
and present a direct proof of the results of the B\"acklund
transformation and the solutions generated. % without twistor framework.
These are due to collaboration with C.~Gilson and J.~Nimmo (Glasgow)
\cite{GHN, GHN2}.

Finally we give an example of noncommutative Ward's conjecture,
reduction of the noncommutative anti-self-dual Yang-Mills 
equation into the noncommutative KdV equation via the 
noncommutative toroidal KdV equation.
The reduced equation actually has integrable-like
properties such as infinite conserved quantities, 
exact N-soliton solutions and so on. These results would lead to 
a kind of classification of noncommutative integrable equations 
{}from a geometrical viewpoint and to applications to the
corresponding physical situations and geometry also.

\section{Noncommutative anti-self-dual Yang-Mills equations}

In this section, we review some aspects of the
noncommutative anti-self-dual Yang-Mills equation and establish notations.
%and present B\"acklund transformations for the noncommutative anti-self-dual 
%Yang-Mills equation in Yang's form and generate 
%noncommutative Atiyah-Ward ansatz solutions in terms of quasideterminants.

\subsection{Noncommutative gauge theories}

Noncommutative spaces are defined
by the noncommutativity of the coordinates:
\begin{eqnarray}
\label{nc_coord}
[x^\mu,x^\nu]=i\theta^{\mu\nu},
\end{eqnarray}
where $\theta^{\mu\nu}$ are real constants
called the {\it noncommutative parameters}. 
The noncommutative parameter is anti-symmetric with respect to $\mu,\nu$:
$\theta^{\nu\mu}=-\theta^{\mu\nu}$ and the rank is even.
This relation looks like the canonical commutation
relation in quantum mechanics
and leads to ``space-space uncertainty relation.''
Hence singularities which exist on commutative spaces
could resolve on noncommutative spaces.
This is one of the prominent features of noncommutative
field theories and yields various new physical objects
such as $U(1)$ instantons.

Noncommutative field theories are given by the exchange of ordinary products
in the commutative field theories for the star-products and
realized as deformed theories {}from the commutative ones. 
The ordering of non-linear terms are determined by
some additional conditions.
% such as the gauge symmetry, 
%and the Lax representation and so on.
The star-product is defined for ordinary fields 
on commutative spaces. 
For Euclidean spaces, it is explicitly given by
\begin{eqnarray}
f\star g(x)&:=&{\mbox{exp}}
\left(\frac{i}{2}\theta^{\mu\nu} \partial^{(x^{\prime})}_\mu
\partial^{(x^{\prime\prime})}_\nu \right)
f(x^\prime)g(x^{\prime\prime})\Big{\vert}_{x^{\prime}
=x^{\prime\prime}=x}\nonumber\\
&=&f(x)g(x)+\frac{i}{2}\theta^{\mu\nu}\partial_\mu f(x)\partial_\nu g(x)
+O (\theta^2),
\label{star}
\end{eqnarray}
where $\del_\mu^{(x^\prime)}:=\del/\del x^{\prime \mu}$
and so on.
This explicit representation is known
as the {\it Moyal product} \cite{Moyal}.
The star-product has associativity:
$f\star(g\star h)=(f\star g)\star h$,
and returns back to the ordinary product
in the commutative limit:  $\theta^{\mu\nu}\rar 0$.
The modification of the product  makes the ordinary
spatial coordinate ``noncommutative,''
that is, $[x^\mu,x^\nu]_\star:=x^\mu\star x^\nu-x^\nu\star x^\mu=i\theta^{\mu\nu}$.

Here are the noncommutative Kadomtsev-Petviashvili (KP) and KdV equations:
\begin{itemize}

 \item NC KP equation in $(2+1)$-dimension 
(typically $[x,y]_\star=i\theta$
or $[t,x]_\star=i\theta$)

\begin{eqnarray}
 \fr{\del u}{\del t}=\frac{1}{4}\fr{\del^3 u}{\del x^3}
+\frac{3}{4}\left(\fr{\del u}{\del x}\star u+u\star \fr{\del u}{\del x}\right)
+\frac{3}{4}\partial_x^{-1} \fr{\del^2 u}{\del y^2}
-\frac{3}{4}\left[u,\partial_x^{-1} \fr{\del u}{\del
y}\right]_\star,
\label{kp}
\end{eqnarray}
where $t$ and $x,y$ are time and spatial coordinates, respectively,
and $\partial_x^{-1}f(x)=\int^x dx^\prime f(x^\prime)$.

\item NC KdV equation in $(1+1)$-dimension ($[t,x]_\star=i\theta$)

\begin{eqnarray}
 \fr{\del u}{\del t}=\frac{1}{4}\fr{\del^3 u}{\del x^3}
+\frac{3}{4}\left(\fr{\del u}{\del x}\star u+u\star \fr{\del u}{\del x}\right).
\label{kdv}
\end{eqnarray}

\end{itemize}
The ordering of variables in non-linear terms is crucial to preserve
some special integrable properties and determined in the Lax
formalism. (For a review, see \cite{Hamanaka_proc2}.)

We note that the fields themselves take c-number values
as usual and the differentiation and the integration for them
are well-defined as usual, for example, $\del_\mu\star \del_\nu
=\del_\mu\del_\nu,~$ and the wedge product of 
$\lambda=\lambda_\mu(x)dx^\mu$ and
$\rho=\rho_\nu(x)dx^\nu$is $\lambda_\mu
\star \rho_\nu dx^\mu\wedge dx^\nu$.

Noncommutative gauge theories are defined in this way
by imposing noncommutative version of the gauge symmetry,
where the gauge transformation is defined as follows:
\begin{eqnarray}
 A_\mu \rightarrow g^{-1}\star A_\mu\star g +g^{-1}\star \del_\mu g,
\end{eqnarray}
where $g$ is an element of the gauge group $G$
(The inverse is assumed to exist in the sense of 
the star product in this paper.)
This is sometimes called the {\it star gauge transformation}.
We note that because of the noncommutativity,
the commutator terms in field strength are always needed
even when the gauge group is abelian in order to preserve
the star gauge symmetry. This $U(1)$ part of the gauge group
actually plays crucial roles in general.
We note that because of the noncommutativity of matrix elements,
cyclic symmetry of traces is broken in general:
\begin{eqnarray}
 \label{trace}
 {\mbox{Tr }}A\star B\neq  {\mbox{Tr }}B\star A.
\end{eqnarray}
Therefore, gauge invariant quantities becomes 
hard to define on noncommutative spaces.

\subsection{Noncommutative anti-self-dual Yang-Mills equations}

Let us consider Yang-Mills theories in 
$4$-dimensional noncommutative spaces whose 
real coordinates of the space are denoted by $(x^0,x^1,x^2,x^3)$,
where the gauge group is $GL(N,\mathbb{C})$.
Here, we follow the convention in \cite{MaWo}.

First, we introduce double null coordinates
of $4$-dimensional space as follows
\begin{eqnarray}
ds^2=2(dzd\tilde{z}-dwd\tilde{w}),
\end{eqnarray}
We can recover various kind of real spaces
by putting the corresponding reality conditions 
on the double null coordinates $z,\zt,w,\wt$ as follows:
\begin{itemize}
 \item Euclidean Space ($\bar{w}=-\wt; \bar{z}=\zt$): 
An example is 
       \begin{eqnarray}
        \left(\begin{array}{cc}\tilde{z}&w\\\tilde{w}&z\end{array}\right)
        =\frac{1}{\sqrt{2}}
         \left(\begin{array}{cc}x^0+ix^1&-(x^2-ix^3)\\x^2+ix^3&x^0-ix^1
               \end{array}\right).
       \end{eqnarray}
 \item Minkowski Space ($\bar{w}=\wt$; $z$ and $\zt$ are real.): 
An example is 
       \begin{eqnarray}
        \left(\begin{array}{cc}\tilde{z}&w\\\tilde{w}&z\end{array}\right)
        =\frac{1}{\sqrt{2}}
         \left(\begin{array}{cc}x^0+x^1&x^2-ix^3\\x^2+ix^3&x^0-x^1
               \end{array}\right).
       \end{eqnarray}
 \item Ultrahyperbolic Space ($\bar{w}=\wt; \bar{z}=\zt$): Example are
       \begin{eqnarray}
        \left(\begin{array}{cc}\tilde{z}&w\\\tilde{w}&z\end{array}\right)
        =\frac{1}{\sqrt{2}}
         \left(\begin{array}{cc}x^0+ix^1&x^2-ix^3\\x^2+ix^3&x^0-ix^1
               \end{array}\right),
~~~{\mbox{or}}~~~z,\zt,w,\wt\in\mathbb{R}.
       \end{eqnarray}
\end{itemize}

The coordinate vectors $\del_z,\del_z.\del_{\wt}, \del_{\zt}$
 form a null tetrad and are represented explicitly as:
 \begin{eqnarray}
 \label{tetrad}
&&  \del_z=\frac{1}{\sqrt{2}}\left(\frac{\del}{\del x^0}+i\frac{\del}{\del
                            x^1}\right),~~~
  \del_{\tilde{z}}=\frac{1}{\sqrt{2}}
  \left(\frac{\del}{\del x^0}-i\frac{\del}{\del
                            x^1}\right),\nn
&&  \del_w=\frac{1}{\sqrt{2}}\left(\frac{\del}{\del x^2}+i\frac{\del}{\del
                            x^3}\right),~~~
  \del_{\tilde{w}}=\frac{1}{\sqrt{2}}
  \left(\frac{\del}{\del x^2}-i\frac{\del}{\del
                            x^3}\right).
\end{eqnarray}

For the Euclidean and ultrahyperbolic signatures, the 
Hodge dual operator $*$ satisfies $*^2=1$ and 
hence the space of 2-forms $\beta$
decomposes into the direct sum of eigenvalues of $*$ 
with eigenvalues $\pm 1$,
that is, self-dual (SD) part ($*\beta=\beta$)
and anti-self-dual (ASD) part ($*\beta=-\beta$). 
{}From now on, we treat these two signatures.

Typical examples of self-dual forms are
\begin{eqnarray}
 \alpha=dw\wedge d{z},~~~\tilde{\alpha}=d\wt\wedge d\zt,
  ~~~\omega=dw\wedge d\wt-dz\wedge d\zt,
  \label{omega}
\end{eqnarray}
and those of anti-self-dual forms are
\begin{eqnarray}
 dw\wedge d{\zt},~~~d\wt\wedge dz,~~~dw\wedge d\wt+dz\wedge d\zt.
\end{eqnarray}

The noncommutative anti-self-dual Yang-Mills equation is derived 
{}from the compatibility condition
of the following linear system:
\begin{eqnarray}
L\star \psi&:=&(D_w-\zeta D_{\tilde{z}})\star \psi=
 \left(\del_w+A_w-\zeta (\del_{\zt}+A_{\tilde{z}})\right)\star \psi(x;\zeta)
= 0,\nn
M\star \psi&:=&(D_z-\zeta D_{\tilde{w}})\star \psi=
  \left(\del_z+A_z-\zeta (\del_{\wt}+A_{\tilde{w}})\right)\star \psi(x;\zeta)
= 0,
\label{lin_asdym}
\end{eqnarray}
where $A_z,A_w,A_{\tilde{z}},A_{\tilde{w}}$
and $D_z,D_w,D_{\tilde{z}},D_{\tilde{w}}$
denote gauge fields and covariant derivatives in the Yang-Mills theory,
respectively. The constant $\zeta\in \mathbb{C}P_1$ is called
the {\it spectral parameter}. 

The compatible condition $[L,M]_\star=0$, gives rise to
a quadratic polynomial of $\zeta$ and each coefficient
yields the following equations:
\begin{eqnarray}
 F_{wz}&=&\del_{w} A_z -\del_{z} A_w+[A_w,A_z]_\star =0,\nn
 F_{\wt\zt}&=&\del_{\tilde{w}} A_{\tilde{z}} -\del_{\tilde{z}} A_{\tilde{w}}
 +[A_{\tilde{w}},A_{\tilde{z}}]_\star =0,\nn
 F_{z\zt}-F_{w\wt}&=&\del_{z} A_{\tilde{z}} -\del_{\tilde{z}} A_{z}
 +\del_{\tilde{w}} A_{w} -\del_{w} A_{\tilde{w}}
 +[A_z,A_{\tilde{z}}]_\star
 -[A_{w},A_{\tilde{w}}]_\star=0,
\label{asdym}
\end{eqnarray}
which are equivalent to
the noncommutative anti-self-dual Yang-Mills equations 
$F_{\mu\nu}=-*F_{\mu\nu}$ in the real representation.

Gauge transformations act on the linear system (\ref{lin_asdym}) as
\begin{eqnarray}
 L\mapsto g^{-1}\star L\star g,~
 M\mapsto g^{-1}\star M\star g,~
 \psi\mapsto g^{-1}\star \psi,~~~g\in G.
\end{eqnarray}

%As in commutative case, if $\Pi$ is a null 2-plane in space-time,
%a tangent bivector $\pi$ associated to $\Pi$ satisfies
%$\pi_{\mu\nu}\pi^{\mu\nu}=0$ and $\pi_{\mu\nu}dx^\mu\wedge dx^\nu$
%is either SD or ASD. 
%For example, vector fields $l=\del_w-\lambda \del_{\zt}$
%and $m=\del_z-\lambda\del_{\wt}$ form a SD bivector $l\wedge m$.
%A 2-plane $\Pi$ is called an {\it $\alpha$-plane} when
%the associated tangent bivector $\pi$ is SD.
%noncommutative anti-self-dual Yang-Mills eqs are equivalent to the condition that
%the connection is flat on the $\alpha$-plane: $F(l,m)=0$.
%(For detailed discussion, see e.g. \cite{MaWo}.)

We note that the solution $\psi$ ($N\times N$ matrix)
of the linear system (\ref{lin_asdym}) 
is not regular at $\zeta=\infty$ because of Liouville's theorem.
(If it is regular, then the gauge fields become flat.)
Hence we have to consider another linear system
on another local patch in $\zeta\in \mathbb{C}P_1$ 
whose coordinate is $\tilde{\zeta}=1/\zeta$ as
\begin{eqnarray}
\tilde{L}\star \tilde{\psi}
&:=&\tilde{\zeta} D_w \star \tilde{\psi}-D_{\tilde{z}}\star \tilde{\psi}=0,\nn
\tilde{M}\star \tilde{\psi}&:=&\tilde{\zeta}D_z\star \tilde{\psi}-D_{\tilde{w}}
\star \tilde{\psi}= 0.
\label{lin_asdym2}
\end{eqnarray}
The compatibility condition of this system
also gives rise to the anti-self-dual Yang-Mills equation.

\subsection{Noncommutative Yang's equations and $J, K$-matrices}

Here we discuss the potential forms of the 
noncommutative anti-self-dual Yang-Mills equations
such as noncommutative $J,K$-matrix formalisms and 
the noncommutative Yang's equation,
which is already presented by e.g. K.~Takasaki \cite{Takasaki}.

Let us first discuss the {\it $J$-matrix formalism}
of the noncommutative anti-self-dual Yang-Mills equation.
The first equation of noncommutative anti-self-dual Yang-Mills equation
(\ref{asdym}) is the compatible condition
of the linear system $D_z\star h=0, D_w\star h=0$,
where $h$ is a $N\times N$ matrix.
Hence we get
\begin{eqnarray}
\label{a}
A_{z}=-(\del_z h)\star h^{-1}, ~~~  A_{w}=-(\del_w h)\star h^{-1}.
\end{eqnarray}
%where $h_z:=\del h/\del z,~h_w:=\del h/\del w$.
Similarly, the second equation of noncommutative anti-self-dual 
Yang-Mills equation (\ref{asdym}) leads to
\begin{eqnarray}
\label{at}
A_{\zt}=-(\del_{\zt}\tilde{h})\star \tilde{h}^{-1}, ~~~
A_{\wt}=-(\del_{\wt}\tilde{h})\star \tilde{h}^{-1},
\end{eqnarray}
where $\tilde{h}$ is also a $N\times N$ matrix.
We note that
$h(x)=\psi(x,\zeta=0), \tilde{h}(x)=\tilde{\psi}(x,\zeta=\infty)$.

By defining a new matrix $J=\tilde{h}^{-1}\star h $, 
the third equation of the noncommutative anti-self-dual Yang-Mills equation 
(\ref{asdym}) becomes the noncommutative Yang's equation 
\begin{eqnarray}
\label{yang}
 \del_z(J^{-1} \star \del_{\zt} J)-\del_w (J^{-1}\star \del_{\wt} J)=0,
\end{eqnarray}
or equivalently,
\begin{eqnarray}
 \del\left(J^{-1}\star \delt J\right)\wedge \omega=0.
\end{eqnarray}
where $\del=dw \del_w +dz\del_z,~\delt=d\wt \del_{\wt} +d\zt\del_{\zt}$
$\omega$ is the same one as in \eqref{omega}. 

Gauge transformations act on $h$ and $\tilde{h}$ as
\begin{eqnarray}
 h\mapsto g^{-1}h,~
 \tilde{h}\mapsto g^{-1}\tilde{h},~~~g\in G.
\end{eqnarray}
Hence the Yang's $J$-matrix is gauge invariant
while the matrices $h$ and $\tilde{h}$ are gauge dependent.
%Gauge fields are obtained from a solution $J$ of the NC Yang's
%equation via a decomposition $J=\tilde{h}^{-1} h$, and \eqref{a} and
%\eqref{at}. 
%The different decompositions correspond to different choices of gauge.
In this paper, we sometimes use the following gauge for $G=GL(2)$:
%for simplicity which realize the parameterization of $J$:
\begin{eqnarray}
&&h_{\scr\mbox{MW}}= \left(\begin{array}{cc}f&0
   \\ e&1\end{array}
   \right),~~~
\tilde{h}_{\scr\mbox{MW}}=\left(\begin{array}{cc}1&g
   \\ 0 &b\end{array}
   \right),\\
&&{\mbox{then}}~~~
 J =\tilde{h}_{\scr\mbox{MW}}^{-1}\star h_{\scr\mbox{MW}}
=\left(\begin{array}{cc} f -g \star b^{-1}\star  e&-g\star  b^{-1}
   \\ b^{-1} \star e &b^{-1}\end{array}
   \right).\nonumber
%,\nn
%\end{eqnarray}
%Then the inverse of $J$ can be easily calculated as 
%\begin{eqnarray} 
\label{M-W}
% J^{-1}=h^{-1}\tilde{h}=
%   \left[\begin{array}{cc}f&0
%   \\ e &1\end{array}
%   \right]^{-1}
% \left[\begin{array}{cc}1&g
%   \\ 0&b\end{array}
%   \right]
%=
%\left[\begin{array}{cc} f^{-1} &f^{-1} g \\
%-e f^{-1} & b-e f^{-1} g
%   \end{array}
%   \right].
\end{eqnarray}
which is called the {\it Mason-Woodhouse gauge}.

There is another potential form of the 
noncommutative anti-self-dual Yang-Mills equation,
known as the {\it $K$-matrix formalism}. 
In the gauge $A_w=A_z=0$,
the third equation of (\ref{asdym})
becomes $\del_z A_{\zt}-\del_w A_{\wt}=0$.
This implies the existence of a potential $K$ such
that $A_{\zt}=\del_{w}K,A_{\wt}=\del_{z}K$.
Then the second equation of (\ref{asdym}) becomes
\begin{eqnarray}
 \del_z\del_{\zt}K -\del_w\del_{\wt}K +[\del_w K, \del_z K]_\star=0.
\end{eqnarray}
Then, 
%Under the gauge $A_w=A_z=0$, 
we get
\begin{eqnarray}
\psi=1+\zeta K+{\cal{O}}(\zeta^2),~~~\tilde{\psi}=J^{-1}+{\cal{O}}(\tilde{\zeta}),
\end{eqnarray}
and $ A_{\wt}=J^{-1}\star \del_{\wt}J= \del_z K,~
 A_{\zt}=J^{-1}\star\del_{\zt}J= \del_w K$.
This gauge is suitable for the discussion of 
the (binary) Darboux transformations for the 
(noncommutative) anti-self-dual Yang-Mills equations \cite{GNO98,GNO00,HSS07}.

\section{Twistor description of noncommutative anti-self-dual Yang-Mills equations}
%Integrable aspects of noncommutative anti-self-dual Yang-Mills equations}

In this section, we construct wide class of exact solutions of 
the noncommutative anti-self-dual Yang-Mills equations
{}from the geometrical viewpoint of the noncommutative twistor theory. 
The noncommutative twistor theory has been developed
by several authors and mathematical foundations are established
\cite{Takasaki, BrMa,Hannabuss,KKO}. 

The twistor theory is based on a correspondence between
(complexified) space-time coordinates $(z,\zt,w,\wt)$
and twistor coordinates $(\lambda,\mu,\zeta)$ which are 
local coordinates of a 3-dimensional complex projective space (twistor space).
The explicit relation is called the {\it incidence relation}, 
and represented as follows:
\begin{eqnarray}
\label{incidence}
\lambda=\zeta w+\zt,~\mu=\zeta z+\wt,
\end{eqnarray}
which implies that for any twistor function $f(\lambda,\mu,\zeta)$,
\begin{eqnarray}
lf(\lambda,\mu,\zeta)&:=&(\del_w-\zeta\del_{\zt})f(\lambda,\mu,\zeta)=0,\nn
mf(\lambda,\mu,\zeta)&:=&(\del_z-\zeta\del_{\wt})f(\lambda,\mu,\zeta)=0.
\label{hol}
\end{eqnarray}

\subsection{Noncommutative Penrose-Ward transformation}

For the anti-self-dual Yang-Mills theory, there is a one-to-one
correspondence between solutions of the anti-self-dual Yang-Mills
equation and holomorphic vector bundles on the twistor space.
The former is given by solutions $\psi,\tilde{\psi}$ of the 
linear systems \eqref{lin_asdym} and \eqref{lin_asdym2}.
The latter is given by patching matrices $P$ of the 
holomorphic vector bundles. The explicit correspondence is
called the {\it Penrose-Ward correspondence}. 

Here we just need the Moyal-deformed Penrose-Ward correspondence
between the anti-self-dual Yang-Mills solution $\psi,\tilde{\psi}$
and the patching matrix $P$. 

{}From given $\psi$ and $\tilde{\psi}$, 
the patching matrix $P$ is constructed as
\begin{eqnarray}
\label{birkhoff}
 P(\zeta w+\zt,\zeta z+\wt,\zeta)=\tilde{\psi}^{-1}(x;\zeta)\star \psi(x;\zeta).
% \psi=\tilde{\psi}\star P,
\end{eqnarray}
(Here we note that $\psi(x;\zeta)$ is regular w.r.t. $\zeta$ around $\zeta=0$
and $\tilde{\psi}(x;\zeta)$ is regular w.r.t. $\tilde{\zeta}$ 
around $\tilde{\zeta}=0$ or equivalently $\zeta=\infty$.)
Conversely, if there exists the factorization
\eqref{birkhoff} into $\psi$ and $\tilde{\psi}$ for a given $P$ 
where $\psi(x;\zeta)$ is regular w.r.t. $\zeta$ around $\zeta=0$
and $\tilde{\psi}(x;\zeta)$ is regular w.r.t. $\tilde{\zeta}$ 
around $\tilde{\zeta}=0$, then the $\psi$ and $\tilde{\psi}$
are solutions of linear systems \eqref{lin_asdym} and \eqref{lin_asdym2} 
for the noncommutative anti-self-dual Yang-Mills equations.
(This factorization problem is called the {\it Riemann-Hilbert problem}
and solved formally \cite{Takasaki}.
Noncommutativity can be introduced into only
two variables $\zeta w+\zt$ and $\zeta z+\wt$. Then $\zeta$ is a commutative
variable and the ways of solving the Riemann-Hilbert problem become
similar to commutative ones. )
%This correspondence is a noncommutaive version of 
%the Penrose-Ward correspondence.

\subsection{Noncommutative Atiyah-Ward ansatz solutions for $G=GL(2)$}

{}From now on, we restrict ourselves to $G=GL(2)$.
For this gauge group, we can take a simple ansatz for
the Patching matrix $P$, which is called the 
{\it Atiyah-Ward ansatz} in the commutative case \cite{AtWa}.
Noncommutative generalization of this ansatz is straightforward 
and actually leads to a solution of the factorization problem.
The $l$-th order noncommutative Atiyah-Ward ansatz 
is specified by the following form of the patching matrix up to
constant matrix actions from both sides ($l=0,1,2,\cdots$):
\begin{eqnarray}
 P_l(x;\zeta)=\left(\begin{array}{cc}0&\zeta^{-l}  
      \\\zeta^{l} &\Delta(x;\zeta)\end{array}
   \right).
\end{eqnarray}
We note that $P$ satisfies eq. \eqref{hol} and 
%implies
%$\Delta(x;\zeta)=(\zeta w+\zt,\zeta z+\wt,\zeta)$,
%or equivalently, $(\del_w-\zeta\del_{\zt})\Delta=0,~(\del_z-\zeta
%\del_{\wt}) \Delta=0$. 
hence, the Laurent expansion of $\Delta$ w.r.t. $\zeta$
\begin{eqnarray}
 \Delta(x;\zeta) = \sum_{i=-\infty}^{\infty}\Delta_i(x) \zeta^{-i},
\end{eqnarray}
gives rise to the following recurrence relations in the coefficients 
as follows
\begin{eqnarray}
\label{rec}
 \frac{\partial \Delta_i}{\partial z}
= \frac{\partial \Delta_{i+1}}{\partial \tilde{w}},~~~
 \frac{\partial \Delta_i}{\partial w}
= \frac{\partial \Delta_{i+1}}{\partial \tilde{z}}.
\end{eqnarray}
%We will soon see that the coefficients $\Delta_i(x)$
%are the scalar functions in the solutions generated by
%the B\"acklund transformations in the previous section.

The wave functions $\psi$ and $\tilde{\psi}$ can be
expanded by $\zeta$ and $\tilde{\zeta}=1/\zeta$, respectively:
\begin{eqnarray}
\label{expansion}
\psi&=&h+{\cal{O}}(\zeta)
=\left(\begin{array}{cc}
h_{11}+\sum_{i=1}^{\infty}a_i\zeta^i&
h_{12}+\sum_{i=1}^{\infty}b_i\zeta^i\\
h_{21}+\sum_{i=1}^{\infty}c_i\zeta^i&
h_{22}+\sum_{i=1}^{\infty}d_i\zeta^i.
\end{array}\right),\nn
\tilde{\psi}&=&\tilde{h}+{\cal{O}}(\tilde{\zeta})
=
\left(\begin{array}{cc}
\tilde{h}_{11}+\sum_{i=1}^{\infty}\at_i\tilde{\zeta}^i&
\tilde{h}_{12}+\sum_{i=1}^{\infty}\bt_i\tilde{\zeta}^i\\
\tilde{h}_{21}+\sum_{i=1}^{\infty}\ct_i\tilde{\zeta}^i&
\tilde{h}_{22}+\sum_{i=1}^{\infty}\dt_i\tilde{\zeta}^i.
\end{array}\right).
\end{eqnarray}

Now let us solve the factorization problem $\tilde{\psi}\star P= \psi$
for the noncommutative Atiyah-Ward ansatz. 
This is concretely written down as
\begin{eqnarray}
\left(\begin{array}{cc} \tilde{\psi}_{11}&\tilde{\psi}_{12}
            \\ \tilde{\psi}_{21}&\tilde{\psi}_{22}\end{array}
   \right)\star
\left(\begin{array}{cc}0&\zeta^{-l} 
      \\ \zeta^l & \Delta(x;\zeta)\end{array}
   \right)
=
\left(\begin{array}{cc}\psi_{11}&\psi_{12}
      \\ \psi_{21}&\psi_{22}\end{array}
   \right),
\end{eqnarray}
that is,
\begin{eqnarray}
\label{splitting1}
&& \tilde{\psi}_{12}\zeta^l=\psi_{11},~~~
\tilde{\psi}_{22}\zeta^l=\psi_{21},\\
&&\tilde{\psi}_{11} \zeta^{-l}
+ \tilde{\psi}_{12}\star \Delta 
=\psi_{12},~~~
\tilde{\psi}_{21} \zeta^{-l} + \tilde{\psi}_{22}\star \Delta
=\psi_{22}.
\label{splitting2}
\end{eqnarray}
{}From Eqs. (\ref{expansion}) and (\ref{splitting1}) 
%the relation between $\psi, \tilde{\psi}$ and $h, \tilde{h}$, 
we find that some entries become polynomials w.r.t. $\zeta$:
\begin{eqnarray}
 \psi_{11}&=&h_{11}+a_1\zeta+a_2\zeta^2+\cdots a_{l-1}\zeta^{l-1}
+\tilde{h}_{12}\zeta^l,\nn
 \psi_{21}&=&h_{21}+b_1\zeta+b_2\zeta^2+\cdots b_{l-1}\zeta^{l-1}
+\tilde{h}_{22}\zeta^l,\nn
 \tilde{\psi}_{12}&=&\tilde{h}_{12}+a_{l-1}\zeta^{-1}
+a_{l-2}\zeta^{-2}+\cdots +a_{1}\zeta^{1-l}
+h_{11} \zeta^{-l},\nn
 \tilde{\psi}_{22}&=&\tilde{h}_{22}+ b_{l-1}\zeta^{-1}
+b_{l-2}\zeta^{-2}+\cdots+b_1 \zeta^{1-l} 
+h_{21} \zeta^{-l},
\end{eqnarray}
and so on. By substituting these relations into eq. \eqref{splitting2},
we get sets of equations for $h$ and $\tilde{h}$ in the coefficients
of $\zeta^{0}, ~\zeta^{-1},~ \cdots, ~\zeta^{-l}$:
\begin{eqnarray}
(h_{11}, a_1,\cdots,a_{l-1},\tilde{h}_{12})
\star D_{l+1}=(-\tilde{h}_{11},0,\cdots,0,h_{12}),\nn
(h_{21}, c_1,\cdots,c_{l-1},\tilde{h}_{22})
\star D_{l+1}=(-\tilde{h}_{21},0,\cdots,0,h_{22}),
\end{eqnarray}
where
\begin{eqnarray}
 D_l:=
\left(
\begin{array}{cccc}
\Delta_0&\Delta_{-1} & \cdots & \Delta_{1-l}\\
\Delta_1 &\Delta_0&\cdots & \Delta_{2-l} \\
\vdots &\vdots &\ddots & \vdots\\
\Delta_{l-1} &\Delta_{l-2} &\cdots &\Delta_0 
\end{array}
\right).
\end{eqnarray}
These linear equations can be solved by taking inverse
matrix of $D_{l+1}$ from right side and 
can be rewritten in terms of quasideterminants 
(For a brief review, see Appendix A.):
\begin{eqnarray}
\label{sol_birkhoff}
 h_{11}&=&h_{12}\star \vert D_{l+1}\vert_{1,l+1}^{-1}
        -\tilde{h}_{11} \star \vert D_{l+1}\vert_{1,1}^{-1},\nn
 h_{21}&=&h_{22}\star \vert D_{l+1}\vert_{1,l+1}^{-1}
        -\tilde{h}_{21} \star \vert D_{l+1}\vert_{1,1}^{-1},\nn
 \tilde{h}_{12}&=&h_{12}\star \vert D_{l+1}\vert_{l+1,l+1}^{-1}
        -\tilde{h}_{1,1} \star \vert D_{l+1}\vert_{l+1,1}^{-1},\nn
 \tilde{h}_{22}&=&h_{22}\star \vert D_{l+1}\vert_{l+1,l+1}^{-1}
        -\tilde{h}_{21} \star \vert D_{l+1}\vert_{l+1,1}^{-1}.
\end{eqnarray}
If we take the Mason-Woodhouse gauge \eqref{M-W}, 
Eq. \eqref{sol_birkhoff} can be solved for $h$ and $\tilde{h}$ 
in terms of quasideterminants of $D_{l+1}$:
\begin{eqnarray}
f=h_{11}&=&-
%(D_l^{-1})_{11}=\vert D_l\vert_{11}^{-1}=
\begin{array}{|cccc|}
\fbox{$\Delta_0$}&\Delta_{-1} & \cdots & \Delta_{-l}\\
\Delta_1 &\Delta_0&\cdots & \Delta_{1-l} \\
\vdots &\vdots &\ddots & \vdots\\
\Delta_{l} &\Delta_{l-1} &\cdots &\Delta_0
\end{array}^{-1},~~~
e=h_{21}=%(D_l^{-1})_{m1}=\vert D_l\vert_{1m}^{-1}=
\begin{array}{|cccc|}
\Delta_0&\Delta_{-1} & \cdots & \fbox{$\Delta_{-l}$}\\
\Delta_1 &\Delta_0&\cdots & \Delta_{1-l} \\
\vdots &\vdots &\ddots & \vdots\\
\Delta_{l} &\Delta_{l-1} &\cdots &\Delta_0
\end{array}^{-1},
\nonumber\\
g=\tilde{h}_{12}%&=&(D_l^{-1})_{1m}=\vert D_l\vert_{m1}^{-1}
&=&-
\begin{array}{|cccc|}
\Delta_0&\Delta_{-1} & \cdots & \Delta_{-l}\\
\Delta_1 &\Delta_0&\cdots & \Delta_{1-l} \\
\vdots &\vdots &\ddots & \vdots\\
\fbox{$\Delta_{l}$} &\Delta_{l-2} &\cdots &\Delta_0 
\end{array}^{-1},~~~
b=\tilde{h}_{22}=%(D_l^{-1})_{mm}=\vert D_l\vert_{mm}^{-1}=
\begin{array}{|cccc|}
\Delta_0&\Delta_{-1} & \cdots & \Delta_{-l}\\
\Delta_1 &\Delta_0&\cdots & \Delta_{1-l} \\
\vdots &\vdots &\ddots & \vdots\\
\Delta_{l} &\Delta_{l-1} &\cdots &\fbox{$\Delta_0$} 
\end{array}^{-1}.
\end{eqnarray}
This is the $l$-th order noncommutative Atiyah-Ward ansatz solution.
For $l=0$, the noncommutative anti-self-dual Yang-Mills equation
becomes a noncommutative linear equation $(\partial_z\partial_{\tilde{z}}
-\partial_w\partial_{\tilde{w}})\Delta_0=0$. (We note that
for the Euclidean space, this is the noncommutative Laplace equation because of
the reality condition $\bar{w}=-\tilde{w}$. The fundamental solutions
leads to noncommutative instanton solutions \cite{NeSc}.) 
The plane wave solutions yields a noncommutative version of 
non-linear plane wave solutions \cite{deVega}.
Other scalar functions $\Delta_i(x)$ is determined explicitly 
by the recurrence relation \eqref{rec} from the solution $\Delta_0(x)$ 
of this linear equation up to integral constants. 
Hence the noncommutative Atiyah-Ward ansatz solutions are exact.

%which coincides exactly with the solutions $R_{l}$  generated 
%by the B\"acklund transformation in the previous section
%with $\epsilon_1=-\epsilon_2=1$.
%That is why we call them the NC Atiyah-Ward ansatz solutions.
%The class of solutions $R_l^\prime$ is also obtained 
%in the same way by starting with the Atiyah-Ward ansatz
%$C_0^{-1}P_lC_0$.

%This just coincides with the generated solutions 
%by the B\"acklund transformation in the previous section !
%This is an origin of the noncommutative Atiyah-Ward solutions.
%\begin{eqnarray}
%(h_{11},a_1,\cdots,a_{l-1},\tilde{h}_{12})\star D_{l+1}
%=(-\tilde{h}_{11},0,\cdots,0,h_{12})
%\left(
%\begin{array}{cccc}
%\Delta_0&\Delta_{-1} & \cdots & \Delta_{-(m-1)}\\
%\Delta_1 &\Delta_0&\cdots & \Delta_{-(m-2)} \\
%\vdots &\vdots &\ddots & \vdots\\
%\Delta_{m-1} &\Delta_{m-2} &\cdots &\Delta_0 
%\end{array}
%\right)
%\star 
%\left(
%\begin{array}{c}
%a_{m-1}\\
%a_{m-2}\\
%\vdots\\
%a_0 (\equiv f)\\ 
%\end{array}
%\right)=-
%\left(
%\begin{array}{c}
%\Delta_1\\
%\Delta_2\\
%\vdots\\
%\Delta_{m}\\ 
%\end{array}
%\right).
%\end{eqnarray}
%By solving this, we can find that $f$ coincides with 
%the one generated by noncommutative CFYG transformation.
%Similar arguments lead to the explicit form of  
%other elements $b, e$ and $g$.

\subsection{B\"acklund transformation for the noncommutative Atiyah-Ward ansatz solutions}

Finally let us discuss an adjoint action for the patching matrices
$\alpha: P_l \mapsto P_{l+1}=A^{-1} P_l A$ in the twistor side,
which leads to a B\"acklund transformation for the noncommutative 
anti-self-dual Yang-Mills equation in the Yang-Mills side.
This is a noncommutative generalization of 
the Corrigan-Fairlie-Yates-Goddard (CFYG) transformation \cite{CFGY,
CMN, MaWo}.

The adjoint action is defined 
by the following two kinds of adjoint actions:
\begin{eqnarray}
 \alpha=\beta\circ\gamma_0,~~~
 \beta: P \mapsto P^{\mbox{\scr{new}}} = B^{-1} P B,~~~
 \gamma_0: P \mapsto P^{\mbox{\scr{new}}} = C_0^{-1} P C_0,
\end{eqnarray}
where
\begin{eqnarray}
 A=BC,~~~
 B=\left(\begin{array}{cc}0&1\\\zeta^{-1}&0\end{array}\right),~~~
 C_0=\left(\begin{array}{cc}0&1\\1&0\end{array}\right).
\end{eqnarray}

In order to find the corresponding transformations in the 
Yang-Mills side, we have to observe how the adjoint actions
act on the matrices $h$ and $\tilde{h}$, or $\psi$ and $\tilde{\psi}$.  
Here we take the Mason-Woodhouse gauge \eqref{M-W}.

We can easily find that the $\gamma_0$-transformation is just
$h\mapsto hC_0, \tilde{h}\mapsto \tilde{h}C_0$
and hence $J=\tilde{h}^{-1}\star h \mapsto C_0^{-1}JC_0$. 
Then, we can read the explicit form of the transformations 
for the variables $b,e,f,g$ in the Mason-Woodhouse gauge \eqref{M-W}.

As for the $\beta$-transformation for $\psi$ and $\tilde{\psi}$,
we have to take a singular gauge transformation 
due to regularity w.r.t. $\zeta$ in the Birkhoff factorization
%$\beta: \psi \mapsto \psi^{\mbox{\scr{new}}}$ and
%$\tilde{\psi} \mapsto \tilde{\psi}^{\mbox{\scr{new}}}$ 
as follows:
\begin{eqnarray}
 \psi^{\mbox{\scr{new}}}=s\star \psi ~B,~~~
 \tilde{\psi}^{\mbox{\scr{new}}}=s\star \tilde{\psi} ~B,
\end{eqnarray}
where the singular gauge transformation is
\begin{eqnarray}
 s=\left(\begin{array}{cc}0&\zeta b^{-1} 
            \\ -f^{-1} &0\end{array}
   \right).
\end{eqnarray}
The explicit calculation gives
\begin{eqnarray}
  \psi^{\mbox{\scr{new}}}
= \left(\begin{array}{cc} b^{-1} \psi_{22}& -\zeta b^{-1}\star \psi_{21} 
            \\ -\zeta^{-1}f^{-1}\star \psi_{12} & f^{-1}\star \psi_{11}
   \end{array}
   \right),
\end{eqnarray}
where $\psi_{ij}$ is the $(i,j)$-th element of $\psi$.
In the $\zeta\rightarrow 0$ limit, this reduces 
to the Mason-Woodhouse gauge:
\begin{eqnarray}
  h^{\mbox{\scr{new}}}
=
\left(\begin{array}{cc}f^{\mbox{\scr{new}}}&0
            \\ e^{\mbox{\scr{new}}} &1\end{array}
   \right)=\left(\begin{array}{cc} b^{-1} & 0 
            \\ - f^{-1} \star j_{12} & 1
   \end{array}
   \right),
   \label{beta-trfed}
\end{eqnarray}
where $\psi=h+j\zeta+{\cal{O}}(\zeta^2)$.

Here we note that the linear systems can be
represented in terms of $b,f,e,g$ as
\begin{eqnarray}
 \label{orig}
 L\star \psi&=&(\del_w-\zeta \del_{\zt})\star \psi+
   \left(\begin{array}{cc}-f_w \star f^{-1}&\zeta g_{\zt}\star b^{-1}
   \\ -e_w\star f^{-1}&\zeta b_{\zt}\star b^{-1}\end{array}
   \right)\star \psi=0,\nn
 M\star \psi&=&(\del_z-\zeta \del_{\wt})\star \psi+
   \left(\begin{array}{cc}- f_z \star f^{-1}&\zeta g_{\wt}\star b^{-1}
   \\ -e_z\star f^{-1}&\zeta b_{\wt}\star b^{-1}\end{array}
   \right)\star \psi=0.
\end{eqnarray}
By picking the first order term of $\zeta$
in the 1-2 component of the first equation, we get
\begin{eqnarray}
 \del_w(f^{-1} \star j_{12})=-f^{-1}\star g_{\zt}\star b^{-1}.
\end{eqnarray}
Hence {}from the 1-1 and 2-1 components of Eq. (\ref{beta-trfed}),
we have
\begin{eqnarray}
 f^{\mbox{\scr{new}}}=b^{-1},~~~
 \del_w e^{\mbox{\scr{new}}}=\del_w(f^{-1}\star j_{12})=-f^{-1}\star
 g_{\zt}\star b^{-1}.
\end{eqnarray}
%which are just parts of the $\beta$-transformation (\ref{new}).
In similar way, we can get the other ones.

%Explicit results are as follows:
%Therefore the $\beta$-transformation (\ref{new})
%can be interpreted as the transformation of the
%patching matrix $F\mapsto B^{-1} F B$ together with the
%singular gauge transformation $s$.

\subsection{Summary and comments}

Here we can reconsider that 
the noncommutative Atiyah-Ward ansatz solutions 
are generated by the two kind of B\"acklund transformation 
from the seed solutions $b=e=f=g=\Delta^{-1}$ without solving 
the Riemann-Hilbert problem.
(The difference of signs in $f,g$ is not essential because they 
can be absorbed into the reflection symmetry $f\mapsto -f,g\mapsto -g$ 
of the noncommutative Yang equation.)  
%{}From this viewpoint, 
Firstly, let us summarize the previous results:

\begin{itemize}
\item $\beta$-transformation \cite{MaWo, Hamanaka06_NPB}: 
\begin{eqnarray}
 \label{new}
   &&e_w^{\mbox{\scriptsize{new}}}=-f^{-1} \star g_{\tilde{z}}\star b^{-1},
   e_z^{\mbox{\scriptsize{new}}}=-f^{-1} \star g_{\tilde{w}} \star b^{-1},\nn
   &&g_{\tilde{z}}^{\mbox{\scriptsize{new}}}=-b^{-1}\star e_w \star f^{-1},
   g_{\tilde{w}}^{\mbox{\scriptsize{new}}}=-b^{-1}\star e_z \star f^{-1},\nn
   &&f^{\mbox{\scriptsize{new}}}=b^{-1},b^{\mbox{\scriptsize{new}}}=f^{-1}.
\end{eqnarray}
%The first four equations %in Eq. (\ref{new}) 
%can be interpreted as integrability conditions
%for the first two equations in (\ref{dYang}). We can easily check
%that the last two equations in  (\ref{dYang})
%are invariant under this transformation.
%This can be considered as a transformation 
%$\beta: J\rightarrow J^{\mbox{\scriptsize{new}}}$. 

\item $\gamma_0$-transformation \cite{GHN}:
\begin{eqnarray}
\left(\begin{array}{cc}
f^{\mbox{\scriptsize{new}}}&g^{\mbox{\scriptsize{new}}}\\
e^{\mbox{\scriptsize{new}}}&b^{\mbox{\scriptsize{new}}}
\end{array}\right)=
\left(\begin{array}{cc}
b&e\\g&f\end{array}\right)^{-1}
=
\left(\begin{array}{cc}
(b-e\star f^{-1} \star g)^{-1}&(g-f \star e^{-1} \star b)^{-1}\\
(e-b \star g^{-1} \star f)^{-1}&(f-g \star b^{-1} \star e)^{-1}\end{array}\right).~~~
\label{gamma_0}
\end{eqnarray}
%This follows {}from the fact that
%the transformation $\gamma_0: J\mapsto 
%J^{\mbox{\scriptsize{new}}}$ is equivalent to 
%the simple conjugation
%\begin{eqnarray}
%\label{CJC}
%$ J^{\mbox{\scriptsize{new}}}=C_0^{-1}JC_0,~
% C_0=\left(\begin{array}{cc}0&1\\1&0\end{array}\right),$
%\end{eqnarray}
%which clearly leaves the noncommutative Yang's equation (\ref{yang}) invariant.
%The relation (\ref{gamma_0}) is derived 
%by comparing elements in this transformation.
%It is a trivial fact that $\gamma_0$-transformation is 
%also involutive.
\end{itemize}
We note that 
both transformations are {\it involutive}, that is,
$\beta\circ\beta$ and $\gamma_0\circ\gamma_0$ are the identity transformations.

Now let us consider the two series of noncommutative Atiyah-Ward ansatz
solutions $R_l$ or $R_l^\prime$ generated by the $\beta\circ\beta$ 
and $\gamma_0\circ\gamma_0$ transformations as follows:
\begin{eqnarray}
\begin{array}{ccccccccc}
R_0&\stackrel{\alpha}{\rightarrow}&R_1&\stackrel{\alpha}{\rightarrow}&
R_2&\stackrel{\alpha}{\rightarrow}&R_3&\rightarrow&\cdots\\
&&&&&&&&\\
~~&{\small{\beta}}\searrow~~&
{\small{\gamma_0}}\updownarrow~~&{\small{\beta}}\searrow~~&
{\small{\gamma_0}}\updownarrow~~&{\small{\beta}}\searrow~~&
{\small{\gamma_0}}\updownarrow~~&\searrow~~&\cdots\\
&&&&&&&&\\
~&~&
R^\prime_1&\stackrel{\alpha^\prime}{\rightarrow}&
R^\prime_2&\stackrel{\alpha^\prime}{\rightarrow}&
R^\prime_3&\rightarrow&\cdots
\end{array}
\end{eqnarray}
where $\alpha=\gamma_0
\circ \beta:R_l \rightarrow R_{l+1}$
and $\alpha^{\prime}=\beta \circ \gamma_0: R^\prime_l \rightarrow
R^\prime_{l+1}$. 
In every solution, the commutative limit leads to $b=f$.
The simplest ansatz $R_0$ and $R_1^\prime$ lead to the so called 
the {\it Corrigan-Fairlie-'t Hooft-Wilczek} 
(CFtHW) ansatz \cite{CoFa} - \cite{Yang}.

Various variables are represented 
in terms of quasideterminants as follows:
\begin{itemize}
 \item Noncommutative Atiyah-Ward ansatz solutions $R_l$

Noncommutative Atiyah-Ward ansatz solutions $R_l$
are represented by the explicit form 
of elements $b_l$, $e_l$, $f_l$, $g_l$ 
as quasideterminants of $(l+1)\times (l+1)$ matrices:
\begin{eqnarray*}
b_l&=&
%(D_m^{-1})_{mm}=\vert D_m\vert_{mm}^{-1}=
\begin{array}{|cccc|}
\Delta_0&\Delta_{-1} & \cdots & \Delta_{-l}\\
\Delta_1 &\Delta_0&\cdots & \Delta_{1-l} \\
\vdots &\vdots &\ddots & \vdots\\
\Delta_{l} &\Delta_{l-1} &\cdots &\fbox{$\Delta_0$} 
\end{array}^{-1},~~~
f_l=
%(D_m^{-1})_{11}=\vert D_m\vert_{11}^{-1}=
\begin{array}{|cccc|}
\fbox{$\Delta_0$}&\Delta_{-1} & \cdots & \Delta_{-l}\\
\Delta_1 &\Delta_0&\cdots & \Delta_{1-l} \\
\vdots &\vdots &\ddots & \vdots\\
\Delta_{l} &\Delta_{l-1} &\cdots &\Delta_0
\end{array}^{-1},\nonumber\\
e_l&=&
%(D_m^{-1})_{m1}=\vert D_m\vert_{1m}^{-1}=
\begin{array}{|cccc|}
\Delta_0&\Delta_{-1} & \cdots & \fbox{$\Delta_{-l}$}\\
\Delta_1 &\Delta_0&\cdots & \Delta_{1-l} \\
\vdots &\vdots &\ddots & \vdots\\
\Delta_{l} &\Delta_{l-1} &\cdots &\Delta_0
\end{array}^{-1},~~~
g_l=
%&=&(D_m^{-1})_{1m}=\vert D_m\vert_{m1}^{-1}
\begin{array}{|cccc|}
\Delta_0&\Delta_{-1} & \cdots & \Delta_{-l}\\
\Delta_1 &\Delta_0&\cdots & \Delta_{1-l} \\
\vdots &\vdots &\ddots & \vdots\\
\fbox{$\Delta_{l}$} &\Delta_{l-1} &\cdots &\Delta_0 
\end{array}^{-1}.
\end{eqnarray*}

\begin{eqnarray*}
\label{J}
J_l=
%\left[
%\begin{array}{cc}
%f'_l-g'_lb_l^{\prime-1}e'_l&-g'_lb_l^{\prime-1}\\
%b_l^{\prime-1}e'_l&b_l^{\prime-1}
%\end{array}\right]=
\begin{pmat}|{||..|}|
\fbox{0}&-1&0&\cdots&0&\fbox{0}\cr\-
1&\Delta_0&\Delta_{-1}&\cdots&\Delta_{1-l}&\Delta_{-l}\cr\-
%\Delta_2&\Delta_1&\Delta_0&\cdots&\Delta_{-l+4}&\Delta_{-l+3}&0\cr
0&\Delta_1&\Delta_0&\cdots&\Delta_{2-l}&\Delta_{1-l}
\cr
\vdots&\vdots&\vdots&\ddots&\vdots&\vdots\cr
0&\Delta_{l-1}&\Delta_{l-2}&\cdots&\Delta_{0}&\Delta_{-1}
\cr\-
\fbox{0}&\Delta_{l}&\Delta_{l-1}&\cdots&\Delta_1&\fbox{$\Delta_{0}$}\cr
\end{pmat},~
J_l^{-1}=
%\left[
%\begin{array}{cc}
%f'_l-g'_lb_l^{\prime-1}e'_l&-g'_lb_l^{\prime-1}\\
%b_l^{\prime-1}e'_l&b_l^{\prime-1}
%\end{array}\right]=
\begin{pmat}|{|..||}|
\fbox{$\Delta_0$}&\Delta_{-1}&\cdots&\Delta_{1-l}&\Delta_{-l}&\fbox{0}\cr\-
%\Delta_2&\Delta_1&\Delta_0&\cdots&\Delta_{-l+4}&\Delta_{-l+3}&0\cr
\Delta_1&\Delta_0&\cdots&\Delta_{2-l}&\Delta_{1-l}&0\cr
\vdots&\vdots&\ddots&\vdots&\vdots&\vdots\cr
\Delta_{l-1}&\Delta_{l-2}&\cdots&\Delta_{0}&\Delta_{-1}&0
\cr\-
\Delta_{l}&\Delta_{l-1}&\cdots&\Delta_{1}&\Delta_{0}&1\cr\-
\fbox{0}&0&\cdots&0&-1&\fbox{$\Delta_{0}$}\cr
\end{pmat}.
\end{eqnarray*}

In the Mason-Woodhouse gauge,
\begin{eqnarray*}
\label{h}
%h_l^{\scr\mbox{MW}}=
h_l=
%\left[
%\begin{array}{cc}
%f'_l-g'_lb_l^{\prime-1}e'_l&-g'_lb_l^{\prime-1}\\
%b_l^{\prime-1}e'_l&b_l^{\prime-1}
%\end{array}\right]=
\begin{pmat}|{||..||}|
\fbox{0}&1&0&\cdots&0&0&\fbox{0}\cr\-
0&\Delta_0&\Delta_{-1}&\cdots&\Delta_{1-l}&\Delta_{-l}&0\cr\-
0&\Delta_1&\Delta_0&\cdots&\Delta_{2-l}&\Delta_{1-l}&0\cr
%0&\Delta_1&\Delta_0&\cdots&\Delta_{2-l}&\Delta_{1-l}\cr
\vdots&\vdots&\vdots&\ddots&\vdots&\vdots&\vdots\cr
0&\Delta_{l-1}&\Delta_{l-2}&\cdots&\Delta_{0}&\Delta_{-1}&0\cr\-
0&\Delta_{l}&\Delta_{l-1}&\cdots&\Delta_{1}&\Delta_{0}&0\cr\-
\fbox{0}&0&0&\cdots&0&1&\fbox{$1$}\cr
\end{pmat},~
%h_l^{\scr\mbox{MW}}^{-1}=
h_l^{-1}=
%\left[
%\begin{array}{cc}
%f'_l-g'_lb_l^{\prime-1}e'_l&-g'_lb_l^{\prime-1}\\
%b_l^{\prime-1}e'_l&b_l^{\prime-1}
%\end{array}\right]=
\begin{pmat}|{|..||}|
\fbox{$\Delta_0$}&\Delta_{-1}&\cdots&\Delta_{1-l}&\Delta_{-l}&\fbox{0}\cr\-
%\Delta_2&\Delta_1&\Delta_0&\cdots&\Delta_{-l+4}&\Delta_{-l+3}&0\cr
\Delta_1&\Delta_0&\cdots&\Delta_{2-l}&\Delta_{1-l}&0\cr
\vdots&\vdots&\ddots&\vdots&\vdots&\vdots\cr
\Delta_{l-1}&\Delta_{l-2}&\cdots&\Delta_{0}&\Delta_{-1}&0
\cr\-
\Delta_{l}&\Delta_{l-1}&\cdots&\Delta_{1}&\Delta_{0}&0\cr\-
\fbox{0}&0&\cdots&0&-1&\fbox{$1$}\cr
\end{pmat}.
\end{eqnarray*}

\begin{eqnarray*}
\label{th}
\tilde{h}_l%^{\scr\mbox{MW}}
=
%\left[
%\begin{array}{cc}
%f'_l-g'_lb_l^{\prime-1}e'_l&-g'_lb_l^{\prime-1}\\
%b_l^{\prime-1}e'_l&b_l^{\prime-1}
%\end{array}\right]=
\begin{pmat}|{||..||}|
\fbox{1}&1&0&\cdots&0&0&\fbox{0}\cr\-
0&\Delta_0&\Delta_{-1}&\cdots&\Delta_{1-l}&\Delta_{-l}&0\cr\-
0&\Delta_1&\Delta_0&\cdots&\Delta_{2-l}&\Delta_{1-l}&0\cr
%0&\Delta_1&\Delta_0&\cdots&\Delta_{2-l}&\Delta_{1-l}\cr
\vdots&\vdots&\vdots&\ddots&\vdots&\vdots&\vdots\cr
0&\Delta_{l-1}&\Delta_{l-2}&\cdots&\Delta_{0}&\Delta_{-1}&0\cr\-
0&\Delta_{l}&\Delta_{l-1}&\cdots&\Delta_{1}&\Delta_{0}&0\cr\-
\fbox{0}&0&0&\cdots&0&1&\fbox{$0$}\cr
\end{pmat},~
\tilde{h}_l^{-1}=
\begin{pmat}|{||..|}|
\fbox{1}&-1&0&\cdots&0&\fbox{0}\cr\-
0&\Delta_0&\Delta_{-1}&\cdots&\Delta_{1-l}&\Delta_{-l}\cr\-
0&\Delta_1&\Delta_0&\cdots&\Delta_{2-l}&\Delta_{1-l}\cr
\vdots&\vdots&\vdots&\ddots&\vdots&\vdots\cr
0&\Delta_{l-1}&\Delta_{l-2}&\cdots&\Delta_0&\Delta_{-1}\cr\-
\fbox{0}&\Delta_{l}&\Delta_{l-1}&\cdots&\Delta_1&\fbox{$\Delta_{0}$}\cr
\end{pmat}.
\end{eqnarray*}

\item Noncommutative Atiyah-Ward ansatz solutions $R^\prime_l$

Noncommutative Atiyah-Ward ansatz solutions $R'_l$
are represented by the explicit form 
of elements $b'_l$, $e'_l$, $f'_l$, $g'_l$
as quasideterminants of $l\times l$ matrices:
\begin{eqnarray*}
b^\prime_l
&=&
%(D_l^{-1})_{11}^{-1}=\vert D_l\vert_{11}=
\begin{array}{|cccc|}
\fbox{$\Delta_0$}&\Delta_{-1} & \cdots & \Delta_{1-l}\\
\Delta_1 &\Delta_0&\cdots & \Delta_{2-l} \\
\vdots &\vdots &\ddots & \vdots\\
\Delta_{l-1} &\Delta_{l-2} &\cdots &\Delta_0 
\end{array}~,~~~
f^\prime_l
=
%(D_l^{-1})_{mm}^{-1}=\vert D_l\vert_{mm}=
\begin{array}{|cccc|}
\Delta_0&\Delta_{-1} & \cdots & \Delta_{1-l}\\
\Delta_1 &\Delta_0&\cdots & \Delta_{2-l} \\
\vdots &\vdots &\ddots & \vdots\\
\Delta_{l-1} &\Delta_{l-2} &\cdots &\fbox{$\Delta_0$}
\end{array}~,\nonumber\\
e^\prime_l
&=&
%(D_l^{-1})_{m1}^{-1}=\vert D_l\vert_{1m}=
\begin{array}{|cccc|}
\Delta_{-1}&\Delta_{-2} & \cdots & \fbox{$\Delta_{-l}$}\\
\Delta_0 &\Delta_{-1}&\cdots & \Delta_{1-l} \\
\vdots &\vdots &\ddots & \vdots\\
\Delta_{l-2} &\Delta_{l-3} &\cdots &\Delta_{-1}
\end{array}~,~~~
g^\prime_l=
%(D_l^{-1})_{1m}^{-1}=\vert D_l\vert_{m1}=
\begin{array}{|cccc|}
\Delta_1&\Delta_{0} & \cdots & \Delta_{2-l}\\
\Delta_2 &\Delta_1&\cdots & \Delta_{3-l} \\
\vdots &\vdots &\ddots & \vdots\\
\fbox{$\Delta_{l}$} &\Delta_{l-1} &\cdots &\Delta_1 
\end{array}~.
\end{eqnarray*}

\begin{eqnarray*}
\label{J_q}
J_l^{\prime}=
%\left[
%\begin{array}{cc}
%f'_l-g'_lb_l^{\prime-1}e'_l&-g'_lb_l^{\prime-1}\\
%b_l^{\prime-1}e'_l&b_l^{\prime-1}
%\end{array}\right]=
\begin{pmat}|{|..||}|
\Delta_0&\Delta_{-1}&\cdots&\Delta_{1-l}&\Delta_{-l}&-1\cr\-
\Delta_1&\Delta_0&\cdots&\Delta_{2-l}&\Delta_{1-l}&0\cr
%\Delta_2&\Delta_1&\Delta_0&\cdots&\Delta_{-l+4}&\Delta_{-l+3}&0\cr
\vdots&\vdots&\ddots&\vdots&\vdots&\vdots\cr
\Delta_{l-1}&\Delta_{l-2}&\cdots&\Delta_0&\Delta_{-1}&0\cr\-
\Delta_{l}&\Delta_{l-1}&\cdots&\Delta_1&\fbox{$\Delta_{0}$}&\fbox{0}\cr\-
1&0&\cdots&0&\fbox{0}&\fbox{0}\cr
\end{pmat},~
%\label{J^-1_q}
J_l^{\prime -1}=
%\left[
%\begin{array}{cc}
%f_l^{\prime-1}&f_l^{\prime-1}g'_l\\
%-e'_lf_l^{\prime-1}&b'_l-e'_lf_l^{\prime-1}g'_l
%\end{array}\right]=
\begin{pmat}|{||..|}|
\fbox{0}&\fbox{0}&0&\cdots&0&1\cr\-
\fbox{0}&\fbox{$\Delta_0$}&\Delta_{-1}&\cdots&\Delta_{1-l}&\Delta_{-l}\cr\-
0&\Delta_1&\Delta_0&\cdots&\Delta_{2-l}&\Delta_{1-l}\cr
\vdots&\vdots&\vdots&\ddots&\vdots&\vdots\cr
0&\Delta_{l-1}&\Delta_{l-2}&\cdots&\Delta_0&\Delta_{-1}\cr\-
-1&\Delta_{l}&\Delta_{l-1}&\cdots&\Delta_1&\Delta_{0}\cr
\end{pmat}.
\end{eqnarray*}

In the Mason-Woodhouse gauge,
\begin{eqnarray*}
h_l^{\prime}=
%\left[
%\begin{array}{cc}
%f'_l-g'_lb_l^{\prime-1}e'_l&-g'_lb_l^{\prime-1}\\
%b_l^{\prime-1}e'_l&b_l^{\prime-1}
%\end{array}\right]=
\begin{pmat}|{|..|}|
\Delta_{l-1}&\Delta_{l-2}&\cdots&\fbox{$\Delta_0$}&\fbox{0}\cr\-
\Delta_{-1}&\Delta_{-2}&\cdots&\fbox{$\Delta_{-l}$}&\fbox{1}\cr\-
\Delta_0&\Delta_{-1}&\cdots&\Delta_{1-l}&0\cr
%&\Delta_1&\cdots&\Delta_{3-l}&\Delta_{2-l}&0\cr
\vdots&\vdots&\ddots&\vdots&\vdots\cr
%0&\Delta_{l-2}&\cdots&\Delta_0&\Delta_{-1}&0\cr\-
\Delta_{l-2}&\Delta_{l-3}&\cdots&\Delta_{-1}&0\cr
\end{pmat},~
h_l^{\prime -1}=
%\left[
%\begin{array}{cc}
%f'_l-g'_lb_l^{\prime-1}e'_l&-g'_lb_l^{\prime-1}\\
%b_l^{\prime-1}e'_l&b_l^{\prime-1}
%\end{array}\right]=
\begin{pmat}|{|..||}|
\fbox{0}&0&\cdots&0&-1&\fbox{0}\cr\-
0&\Delta_0&\cdots&\Delta_{2-l}&\Delta_{1-l}&0\cr
%\Delta_2&\Delta_1&\Delta_0&\cdots&\Delta_{-l+4}&\Delta_{-l+3}&0\cr
\vdots&\vdots&\ddots&\vdots&\vdots&\vdots\cr
0&\Delta_{l-2}&\cdots&\Delta_0&\Delta_{-1}&0\cr\-
1&\Delta_{l-1}&\cdots&\Delta_1&\Delta_{0}&0\cr\-
\fbox{0}&\Delta_{-1}&\cdots&\Delta_{1-l}&\Delta_{-l}&\fbox{1}\cr
\end{pmat},~\\
\tilde{h}_l^{\prime}=
%\left[
%\begin{array}{cc}
%f'_l-g'_lb_l^{\prime-1}e'_l&-g'_lb_l^{\prime-1}\\
%b_l^{\prime-1}e'_l&b_l^{\prime-1}
%\end{array}\right]=
\begin{pmat}|{|..|}|
0&\Delta_1&\cdots&\Delta_{3-l}&\Delta_{2-l}\cr
%&\Delta_1&\cdots&\Delta_{3-l}&\Delta_{2-l}&0\cr
\vdots&\vdots&\ddots&\vdots&\vdots\cr
%0&\Delta_{l-2}&\cdots&\Delta_0&\Delta_{-1}&0\cr\-
0&\Delta_{l-1}&\cdots&\Delta_{1}&\Delta_{0}\cr\-
\fbox{1}&\fbox{$\Delta_l$}&\cdots&\Delta_{2}&\Delta_1\cr\-
\fbox{0}&\fbox{$\Delta_0$}&\cdots&\Delta_{2-l}&\Delta_{1-l}\cr
\end{pmat},~
\tilde{h}_l^{\prime -1}=
%\left[
%\begin{array}{cc}
%f'_l-g'_lb_l^{\prime-1}e'_l&-g'_lb_l^{\prime-1}\\
%b_l^{\prime-1}e'_l&b_l^{\prime-1}
%\end{array}\right]=
\begin{pmat}|{|..||}|
\fbox{1}&\Delta_l&\cdots&\Delta_2&\Delta_1&\fbox{0}\cr\-
0&\Delta_0&\cdots&\Delta_{2-l}&\Delta_{1-l}&0\cr\-
0&\Delta_1&\cdots&\Delta_{3-l}&\Delta_{2-l}&0\cr
\vdots&\vdots&\ddots&\vdots&\vdots&\vdots\cr
%0&\Delta_{l-2}&\cdots&\Delta_0&\Delta_{-1}&0\cr\-
0&\Delta_{l-1}&\cdots&\Delta_1&\Delta_{0}&1\cr\-
\fbox{0}&0&\cdots&0&-1&\fbox{0}\cr
\end{pmat},~
\end{eqnarray*}
where
\begin{align*}
\begin{vmatrix}
a_{11}&a_{12}&a_{13}&a_{14}\\
a_{21}&a_{22}&a_{23}&a_{24}\\
a_{31}&a_{32}&\fbox{$a_{33}$}&\fbox{$a_{34}$}\\
a_{41}&a_{42}&\fbox{$a_{43}$}&\fbox{$a_{44}$}\\
\end{vmatrix}&
:=
\begin{bmatrix}
\begin{vmatrix}
a_{11}&a_{12}&a_{13}\\
a_{21}&a_{22}&a_{23}\\
a_{31}&a_{32}&\fbox{$a_{33}$}
\end{vmatrix}&\begin{vmatrix}
a_{11}&a_{12}&a_{14}\\
a_{21}&a_{22}&a_{24}\\
a_{31}&a_{32}&\fbox{$a_{34}$}
\end{vmatrix}
\\
\begin{vmatrix}
a_{11}&a_{12}&a_{13}\\
a_{21}&a_{22}&a_{23}\\
a_{41}&a_{42}&\fbox{$a_{43}$}\\
\end{vmatrix}&
\begin{vmatrix}
a_{11}&a_{12}&a_{14}\\
a_{21}&a_{22}&a_{24}\\
a_{41}&a_{42}&\fbox{$a_{44}$}\\
\end{vmatrix}
\end{bmatrix}.
\end{align*}
%Here lower case letters denote single entries, upper case letters
%denote matrices of compatible dimensions and Greek letters are
%scalars (i.e.{} commute with everything). 

\end{itemize}

Because $J$ is gauge invariant, this shows that
the present B\"acklund transformation is not
just a gauge transformation but a non-trivial one.

The proof of these results can be made directly 
by using identities of quasideterminants only, such as,
noncommutative Jacobi identity, homological relations,
and Gilson-Nimmo's derivative formula \cite{GHN, GHN2}.
(For the compact representations of $J$, see especially 
Appendix A in \cite{GHN}.)
This implies that {\it noncommutative B\"acklund transformations
are identities of quasideterminants}.

%The explicit form of the solutions 
%can be represented in terms of quasideterminants
%whose elements $\Delta_r$ ($r=-l+1,-l+2,\cdots,l-1$) satisfy
%\begin{eqnarray}
%\label{chasing}
% \frac{\partial \Delta_r}{\partial z}
%= \frac{\partial \Delta_{r+1}}{\partial \tilde{w}},~~~
% \frac{\partial \Delta_r}{\partial w}
%= \frac{\partial \Delta_{r+1}}{\partial \tilde{z}},~~~
%-l+1\leq r\leq l-2~~~(l\geq 2),
%\end{eqnarray}
%which implies that every element $\Delta_r$ is a solution of
%the noncommutative linear equation $(\partial_z\partial_{\tilde{z}}
%-\partial_w\partial_{\tilde{w}})\Delta_r=0$.

\section{Noncommutative Ward's Conjecture}

Here we briefly discuss reductions of the 
noncommutative anti-self-dual Yang-Mills equation
into lower-dimensional noncommutative integrable equations such as the 
noncommutative KdV equation. %let us summarize the strategy for
%the reductions. of noncommutative anti-self-dual Yang-Mills equation 
%into lower-dimensions.
The reductions are specified by a choice of 
gauge group, symmetry, gauge fixing and so on. 
Gauge groups are in general $GL(N)$.
We have to take the $U(1)$ part of the gauge group 
into account in noncommutative case.
%A choice of symmetry reduces
%noncommutative anti-self-dual Yang-Mills equations to simple forms.
The noncommutativity
in the reduced directions is assumed to be eliminated
because of compatibility with the symmetry.
(Hence within the reduced directions, 
the symmetry is the same as commutative one.)
%The choice of gauge fixing is non-trivial.
The residual gauge symmetry sometimes shows equivalence of a few
reductions. 
%Here we neglect constants of integrations in the process of reductions.

%\subsection{Reduction to the noncommutative KdV Equation}

\bigskip
Here, we present non-trivial reductions of the 
noncommutative anti-self-dual Yang-Mills equation with $G=GL(2)$ 
to the noncommutative KdV equation via a $(2+1)$-dimensional
integrable equation.

Let us start with the standard anti-self-dual Yang-Mills 
equation \eqref{asdym} with $G=GL(2,\mathbb{C})$)
and impose the following translational invariance:
\begin{eqnarray}
 Y=\del_{\tilde{z}},
\end{eqnarray}
and put the following non-trivial 
reduction conditions for the gauge fields:
\begin{eqnarray*}
&&A_{\tilde{w}}=O,~
A_{\tilde{z}}=\left(\begin{array}{cc}0&0\\1&0\end{array}\right),~
A_{w}=\left(\begin{array}{cc}q&~-1\\q_w+q \star q
&~-q\end{array}\right),~\nn
&&A_{z}=\left(\begin{array}{cc}
\displaystyle (1/2)q_{w\tilde{w}}+
q_{\tilde{w}} \star  q+\alpha
&-q_{\tilde{w}}\\
\phi&
\displaystyle -(1/2)q_{w\tilde{w}}
-q \star q_{\tilde{w}}+\alpha
            \end{array}\right),
\end{eqnarray*}
where
\begin{eqnarray*}
\alpha&=&\del_w^{-1}[q_w, q_{\tilde{w}}]_\star,~~~~~~
\del_w^{-1} f(w):=\int^w dw^\prime f(w^\prime),~
\left\{A,B\right\}_\star :=A \star B +B \star A,\nn
\phi&=&\displaystyle
-q_z+\frac{1}{2}q_{ww\tilde{w}}
+\frac{1}{2}\left\{q,q_{w\tilde{w}}\right\}_\star
+\frac{1}{2} \left\{q_{w}, q_{\tilde{w}}\right\}_\star
+q \star q_{\tilde{w}} \star q
+[q,\del_w^{-1}[q_w, q_{\tilde{w}}]_\star]_\star.\nonumber
\end{eqnarray*}
Then we get a noncommutative version of the 
toroidal KdV equation \cite{HaTo}
by identifying $2q_w =u$:
\begin{eqnarray}
 u_z=\displaystyle\frac{1}{4}u_{ww\tilde{w}}
+\frac{1}{2}\left\{u, u_{\tilde{w}}\right\}_\star
+\frac{1}{4} \left\{u_{\tilde{w}}, \del_w^{-1} u_{\tilde{w}}
             \right\}_\star
+\frac{1}{4}\del_w^{-1}[u,
\del_w^{-1}[u, \del_w^{-1} u_{\tilde{w}}]_\star]_\star.
\label{bcs} 
\end{eqnarray}
This equation has hierarchy and N-soliton 
solutions in terms of quasideterminants of the Wronskian \cite{Hamanaka06_JHEP}.
We note that under the ultrahyperbolic signature 
$(++--)$, all remaining coordinates among $z,w,\tilde{w}$
can be set to be real \cite{MaWo}. 

If we take further reduction $\del_w=\del_{\tilde{w}}$,
that is, dimensional reduction to the $X=\del_w-\del_{\tilde{w}}$
direction, then the reduced equation coincides with 
the noncommutative KdV equation:
\begin{eqnarray}
 \dot{u}=\displaystyle
\frac{1}{4}u^{\prime\prime\prime}
+\frac{3}{4}\left(u^{\prime} \star u+u \star  u^{\prime}\right).
\end{eqnarray}
where $(t,x)\equiv (z,w+\tilde{w})$
and $\dot{f}:=\del f/\del t,~f^\prime:=\del f/\del x$.
We note that the gauge group is not 
$SL(2)$ but $GL(2)$
because $A_z$ is not traceless. This implies that
the $U(1)$ part of the gauge group plays a crucial role
in the reduction process also.

This noncommutative KdV equation has been studied by several authors and
proved to possess infinite conserved quantities
\cite{DiMH} in terms of Strachan's products \cite{Strachan}
and exact multi-soliton solutions in terms of 
quasideterminants \cite{EGR, Hamanaka06_JHEP}.
(See also \cite{Paniak}.)

\section{Conclusion and Discussion}

In this paper, we have presented the B\"acklund transformations for the
noncommutative anti-self-dual Yang-Mills
equation with $G=GL(2)$ and constructed 
a series of the exact noncommutative Atiyah-Ward ansatz solutions 
in terms of quasideterminants.
%We have also discussed the origin of
%this transformation in the framework of noncommutative twistor theory.

The quasideterminants play important roles 
in the construction of noncommutative soliton solutions 
not only for the noncommutative anti-self-dual Yang-Mills
equation, but also various lower-dimensional
noncommutative integrable equations 
\cite{DiMH2}-\cite{SaPe}.
%\cite{DiMHGiMa, GiNi, GNO, GNS, GNS2,
%GoVe, HaHa, Hassan, LiNi, LiNi2, LNT, Nimmo, ReRu, HSS05, SaPe}.

Such common properties have been revealed
in the study of the noncommutative extension;
however, even within the commutative limit, 
it gives us a new insight.
Various properties and identities of the 
quasideterminants are actually very useful and suitable
for the noncommutative soliton theory.
Surprisingly, obtaining a proof by using the quasideterminants 
is sometimes easier than achieving the same end 
by using the commutative determinants!
This suggests that the quasideterminants might be
more essential than the usual determinants in the 
context of soliton theories. (even within the commutative limit!)
In Sato's theory of solitons, the Pl\"ucker relations 
of the Wronskian play crucial roles.
The present results would suggest the possibility of 
both noncommutative extension and higher-dimensional 
extension of his theory.
It might be time to reconsider a formulation of Sato's theory
of (noncommutative) anti-self-dual Yang-Mills equations
from the viewpoint of quasideterminants.
(For commutative discussions, see e.g. \cite{Takasaki_CMP, MOS})

\section*{Acknowledgements}

The author would like to thank the organizers of
the RIMS International Conference on
Geometry related to Integrable Systems,
25 - 28 September, 2007 in Kyoto, and 
the COE workshop on Noncommutative Geometry and
Physics, 26 February - 3 March, 2008 at Shonan Village Center, Japan
for the invitation to present this work
and for their hospitality.
He is grateful to C.~Gilson and J.~Nimmo
for a fruitful collaboration leading to the completion of this work, 
and to 
T.~Asakawa
A.~Dimakis, 
I.~Kishimoto, 
O.~Lechtenfeld, 
L.~Mason,
F.~M\"uller-Hoissen, 
Y.~Ohta and
K.~Takasaki for useful comments.
Thanks are due to the organizers and audiences during the workshops 
YITP-W-09-04 on ``QFT 2009'' for their hospitality and discussion,
respectively. 
This work was partially supported by 
%JSPS Research Fellowships for Young Scientists (\#0310363),
%the Nishina Memorial Foundation, 
%the Inoue Foundation
the Daiko Foundation,
%the _Nagoya University Foundation, 
the Showa Public-Reward Foundation
and the Toyoaki Scholarship Foundation.

\begin{appendix}

\section{Brief Review of Quasideterminants}

In this section, we make a brief introduction
of quasideterminants introduced by Gelfand and Retakh
in 1991 \cite{GeRe} and present a few properties
of them which play important roles in section 4.
A good survey is e.g. \cite{GGRW} and 
relation between quasideterminants and 
noncommutative symmetric functions 
is summarized in e.g. \cite{GKLLRT}. (See also, \cite{KrLe, Suzuki})

Quasideterminants are not just a noncommutative 
generalization of usual commutative determinants 
but rather related to inverse matrices.

Let $A=(a_{ij})$ be a $n\times n$ matrix and 
$B=(b_{ij})$ be the inverse matrix of $A$.
Here all matrix elements are supposed to 
belong to a (noncommutative) ring with an associative product.
This general noncommutative situation includes 
the Moyal or noncommutative deformation which we discuss in the main sections.

Quasideterminants of $A$ are defined formally
as the inverse of the elements of $B=A^{-1}$:
\begin{eqnarray}
 \vert A \vert_{ij}:=b_{ji}^{-1}.
\end{eqnarray}
In the commutative limit, this is reduced to
\begin{eqnarray}
 \vert A \vert_{ij} \longrightarrow
  (-1)^{i+j}\frac{\det A}{\det \tilde{A}^{ij}},
\label{limit}
\end{eqnarray}
where $\tilde{A}^{ij}$ is the matrix obtained {}from $A$
deleting the $i$-th row and the $j$-th column.

We can write down more explicit form of quasideterminants.
In order to see it, let us recall the following formula
for a square matrix:
\begin{eqnarray}
 \left(
 \begin{array}{cc}
  A&B \\C&D
 \end{array}
 \right)^{-1}
=\left(\begin{array}{cc}
% (A-BD^{-1}C)^{-1}&(C-DB^{-1}A)^{-1}\\(B-AC^{-1}D)^{-1}&(D-CA^{-1}B)^{-1}
 (A-BD^{-1}C)^{-1}&-A^{-1}B(D-CA^{-1}B)^{-1}\\-(D-CA^{-1}B)^{-1}CA^{-1}
&(D-CA^{-1}B)^{-1}
\end{array}\right),
\end{eqnarray}
where $A$ and $D$ are square matrices, and all inverses
are supposed to exist. We note that any matrix can be decomposed
as a $2\times 2$ matrix by block decomposition
where the diagonal parts are square matrices,
and the above formula can be applied to the decomposed
$2\times 2$ matrix.
So the explicit forms of quasideterminants 
are given iteratively by the following formula:
\begin{eqnarray}
 \vert A \vert_{ij}&=&a_{ij}-\sum_{i^\prime (\neq i), j^\prime (\neq j)}
  a_{ii^\prime} ((\tilde{A}^{ij})^{-1})_{i^\prime j^\prime} a_{j^\prime
  j}\nonumber\\
 &=&a_{ij}-\sum_{i^\prime (\neq i), j^\prime (\neq j)}
  a_{ii^\prime} (\vert \tilde{A}^{ij}\vert_{j^\prime i^\prime })^{-1}
  a_{j^\prime j}.
\end{eqnarray}

It is sometimes convenient to represent the quasideterminant
as follows:
\begin{eqnarray}
 \vert A\vert_{ij}=
  \begin{array}{|ccccc|}
   a_{11}&\cdots &a_{1j} & \cdots& a_{1n}\\
   \vdots & & \vdots & & \vdots\\
   a_{i1}&~ & {\fbox{$a_{ij}$}}& ~& a_{in}\\
   \vdots & & \vdots & & \vdots\\
   a_{n1}& \cdots & a_{nj}&\cdots & a_{nn}
  \end{array}~.
\end{eqnarray}

Examples of quasideterminants are,
for a $1\times 1$ matrix $A=a$
 \begin{eqnarray}
  \vert A \vert  = a,
 \end{eqnarray}
and 
for a $2\times 2$ matrix $A=(a_{ij})$
 \begin{eqnarray}
  \vert A \vert_{11}=
   \begin{array}{|cc|}
   \fbox{$a_{11}$} &a_{12} \\a_{21}&a_{22}
   \end{array}
 =a_{11}-a_{12}a_{22}^{-1}a_{21},~~~
  \vert A \vert_{12}=
   \begin{array}{|cc|}
   a_{11} & \fbox{$a_{12}$} \\a_{21}&a_{22}
   \end{array}
 =a_{12}-a_{11}a_{21}^{-1}a_{22},\nonumber\\
  \vert A \vert_{21}=
   \begin{array}{|cc|}
   a_{11} &a_{12} \\ \fbox{$a_{21}$}&a_{22}
   \end{array}
 =a_{21}-a_{22}a_{12}^{-1}a_{11},~~~
  \vert A \vert_{22}=
   \begin{array}{|cc|}
   a_{11} & a_{12} \\a_{21}&\fbox{$a_{22}$}
   \end{array}
 =a_{22}-a_{21}a_{11}^{-1}a_{12}, 
 \end{eqnarray}
 and for a $3\times 3$ matrix $A=(a_{ij})$
  \begin{eqnarray}
  \vert A \vert_{11}
   &=&
   \begin{array}{|ccc|}
   \fbox{$a_{11}$} &a_{12} &a_{13}\\ a_{21}&a_{22}&a_{23}\\a_{31}&a_{32}&a_{33}
   \end{array}
=a_{11}-(a_{12}, a_{13})\left(
\begin{array}{cc}a_{22} & a_{23} \\a_{32}&a_{33}\end{array}\right)^{-1}
\left(
\begin{array}{c}a_{21} \\a_{31}\end{array}
\right)
\nonumber\\
%  &=&a_{11}-a_{12} ((\tilde{A}^{11})^{-1})_{22}  a_{21}
%           -a_{12} ((\tilde{A}^{11})^{-1})_{23}  a_{31}
%           -a_{13} ((\tilde{A}^{11})^{-1})_{32}  a_{21}
%           -a_{13} ((\tilde{A}^{11})^{-1})_{33}  a_{31}\nonumber\\
%  &=&a_{11}-a_{12} \vert \tilde{A}^{11} \vert_{22}  a_{21}
%           -a_{12} \vert \tilde{A}^{11} \vert_{32}  a_{31}
%           -a_{13} \vert \tilde{A}^{11} \vert_{23}  a_{21}
%           -a_{13} \vert \tilde{A}^{11} \vert_{33}  a_{31},\nonumber\\
  &=&a_{11}-a_{12}~ \begin{array}{|cc|}
                   \fbox{$a_{22}$} & a_{23} \\a_{32}&a_{33}
                   \end{array}^{-1}  a_{21}
           -a_{12}~ \begin{array}{|cc|}
                   a_{22} & a_{23} \\\fbox{$a_{32}$}&a_{33}
                   \end{array}^{-1}  a_{31}      \nonumber\\
&&~~~~    -a_{13}~ \begin{array}{|cc|}
                   a_{22} &\fbox{$a_{23}$} \\a_{32}&a_{33}
                                         \end{array}^{-1}  a_{21}
           -a_{13}~ \begin{array}{|cc|}
                   a_{22} & a_{23} \\a_{32}&\fbox{$a_{33}$}
                   \end{array}^{-1} a_{31},
 \end{eqnarray}
and so on.

Quasideterminants have various interesting properties
similar to those of determinants. Among them,
the following ones play important roles in this paper. 
In the block matrices given in these results, 
lower case letters denote single entries and upper case
letters denote matrices
 of compatible dimensions so that the overall matrix is square. 
(By using boxes, it becomes easier to calculate 
various identities. Such calculations are fully 
presented in e.g. \cite{GiNi, Nimmo}.)

\begin{itemize}

 \item Noncommutative Jacobi identity \cite{GeRe, GiNi}
%A simple and useful special case of the noncommutative Sylvester's Theorem
%       \cite{GeRe} is
\begin{equation*}\label{nc syl}
    \begin{vmatrix}
      A&B&C\\
      D&f&g\\
      E&h&\fbox{$i$}
    \end{vmatrix}=
    \begin{vmatrix}
      A&C\\
      E&\fbox{$i$}
    \end{vmatrix}-
    \begin{vmatrix}
      A&B\\
      E&\fbox{$h$}
    \end{vmatrix}
    \begin{vmatrix}
      A&B\\
      D&\fbox{$f$}
    \end{vmatrix}^{-1}
    \begin{vmatrix}
      A&C\\
      D&\fbox{$g$}
    \end{vmatrix}.
\end{equation*}

\item Homological relations \cite{GeRe}
\begin{eqnarray*}\label{row hom}
    \begin{vmatrix}
      A&B&C\\
      D&f&g\\
      E&\fbox{$h$}&i
    \end{vmatrix}
=  \begin{vmatrix}
      A&B&C\\
      D&f&g\\
      E&h&\fbox{$i$}
    \end{vmatrix}
    \begin{vmatrix}
      A&B&C\\
      D&f&g\\
      0&\fbox{0}&1
    \end{vmatrix},~~~
\label{col hom}
    \begin{vmatrix}
      A&B&C\\
      D&f&\fbox{$g$}\\
      E&h&i
    \end{vmatrix}
=      \begin{vmatrix}
      A&B&0\\
      D&f&\fbox{0}\\
      E&h&1
    \end{vmatrix}
    \begin{vmatrix}
      A&B&C\\
      D&f&g\\
      E&h&\fbox{$i$}
    \end{vmatrix}
\end{eqnarray*}

\item {\it Gilson-Nimmo's derivative formula} \cite{GiNi} 

\begin{align*}
    \begin{vmatrix}
    A&B\\
    C&\fbox{$d$}
    \end{vmatrix}'
 \label{col diff} &=
    \begin{vmatrix}
    A&B'\\
    C&\fbox{$d'$}
    \end{vmatrix}
    +\sum_{k=1}^n
    \begin{vmatrix}
    A&(A_k)'\\
    C&\fbox{$(C_k)'$}
    \end{vmatrix}
    \begin{vmatrix}
    A&B\\
    e_k^t&\fbox{$0$}
    \end{vmatrix},
\end{align*}
where $A_k$ is the $k$th column of a matrix $A$ and  
$e_k$ is the column $n$-vector $(\delta_{ik})$ (i.e.\ 1 in the
$k$th row and 0 elsewhere). 

\end{itemize}

\end{appendix}

\end{document}